\documentclass[]{aastex631}

\shorttitle{Unresolved binaries and multiples in open clusters}
\shortauthors{Malofeeva et al.}

\graphicspath{{./}{figures/}}

\begin{document}

\title{Unresolved Binaries and Multiples in the Intermediate Mass Range in open clusters: Pleiades, Alpha Per, Praesepe, and NGC 1039.}

\correspondingauthor{Giovanni Carraro}
\email{giovanni.carraro@unipd.it}

\author{Alina A. Malofeeva}
\affiliation{Ural Federal University \\
19 Mira Street, 620002 Ekaterinburg, Russia}

\author{Varvara O. Mikhnevich}
\affiliation{Ural Federal University \\
19 Mira Street, 620002 Ekaterinburg, Russia}

\author[0000-0002-0155-9434]{Giovanni Carraro}
\affiliation{Dipartimento di Fisica e Astronomia, Universita'  di Padova \\
Vicolo Osservatorio 3, I35122, Padova, Italy}

\author[0000-0001-8669-803X]{Anton F. Seleznev}
\affiliation{Ural Federal University \\
19 Mira Street, 620002 Ekaterinburg, Russia}
 
\begin{abstract}

In this study, we continue our project to search for unresolved binary and multiple systems in open clusters exploiting the photometric diagram (H-W2)-W1 vs W2-(BP-K) firstly introduced in \citet{Malofeeva+2022}.
In particular, here we estimate the binary and multiple star ratios and the distribution of the component mass ratio $q$ in the Galactic clusters Alpha Persei, Praesepe, and NGC 1039.
We have modified the procedure outlined in our first study \citep{Malofeeva+2022} making star counts automatic and by introducing bootstrapping for error estimation.
Basing on this, we re-investigated the Pleiades star cluster in the same mass range as in our previous work and corrected an inaccuracy in the mass ratio $q$ distribution.
The binary and multiple star ratio in the four clusters is then found to lie between 0.45$\pm$0.03 and 0.73$\pm$0.03.
On the other hand, the ratio of systems with multiplicity more than 2 is between 0.06$\pm$0.01 and 0.09$\pm$0.02.
The distribution of the component mass ratio $q$ is well fitted with a Gaussian having the mode between 0.22$\pm$0.04 and 0.52$\pm$0.01 and the dispersion between 0.10$\pm$0.02 and 0.35$\pm$0.07.
All clusters show a large number of the very low-mass secondary components in the binary systems with primary components below 0.5 $M_{\odot}$.

\end{abstract}


\section{Introduction} \label{sec:intro}

The first evidences that open star clusters must contain a significant number of unresolved binary systems came out as early as the first half of the last century \citep{Haffner&Heckmann1937}.
Since then, a large number of studies have been carried out which demonstrated that the fraction of unresolved binary stars in open clusters usually exceeds 30\% \citep{Bonifazi+1990, Khalaj&Baumgardt2013, Sarro+2014, Sheikhi+2016, Li+2017}.
A comprehensive review of the topic is given in \citet{Duchene&Kraus2013}.

As for the fraction of binary systems, it is widely believed that it increases at increasing the mass of the primary component. Opposite situations however have been reported:  \citet{Patience+2002} for instance found an increase of the fraction of binaries towards lower masses (hence redder colors) in the Alpha Persei and Praesepe clusters.

In the literature various expressions can be encountered for the distribution of the mass ratio of the components of binary systems $q=M_2/M_1$, where $M_1$ is the mass of the primary component, and $M_2$ is the secondary.
Most researchers use a flat distribution (all ratios have the same probability).
There are however contrasting opinions on whether the q distribution has a maximum near $q=1$.
\citet{Duquennoy&Mayor1991} in fact report that there is no such maximum, whereas \citet{Fisher+2005} and \citet{Maxted+2008} claim that such a maximum exists both for field stars and for stars in young star clusters.
Also \citet{Raghavan+2010} support this point of view.
Along the same vein, \citet{El-Badry+2019} presented the strong evidence on the sharp peak at $q=1$ for field stars at the base of Gaia data.
\citet{Reggiani&Meyer2013} (for field stars) and \citet{Kouwenhoven+2009} (for field stars and clusters) first proposed a power law for the $q$ distribution.
\citet{Kouwenhoven+2009} also suggested a Gaussian distribution for $q$.
Finally, according to \citet{Patience+2002}, the q distribution is different for different mass intervals.

Thus, the currently available data on the properties of the population of unresolved binary stars in clusters are clearly incomplete and lead to 
contradictory results.

It should also be noted that data on the q distribution are usually obtained for sufficiently large values of this ratio, since the position of unresolved binary stars with small $q$ on widely used photometric diagrams in the visible wavelengths differs very little from the position of single stars.\\

\noindent
The identification and classification of unresolved binary systems in star clusters based on multi-color stellar photometry have been the subject of a number of relatively recent papers.
\citet{Malkov+2010} exploit the Gaia photometric system  and suggest to use photometric bands in the ultraviolet range to search for unresolved binary systems \citep{Malkov+2011} but,
unfortunately, there are currently no available deep enough photometric datasets in the ultraviolet range.
\citet{BardalezGagliuffi+2014} have proposed  methods to identify the unresolved binaries among the very low-mass stars and brown dwarfs using spectroscopic observations.
\citep{Geissler+2011} used the cross-comparison of data from the SDSS and 2MASS surveys for this goal.
Unfortunately, the use of only JHK filters from the 2MASS survey is insufficient for the reliable detection of unresolved binaries with a small $q$ value.

\citet{Borodina&Kovaleva2020} proposed an original method for studying the population of unresolved binary stars based on Gaia photometry, but, again, the use of photometry only in the visible domain does not allow one to detect unresolved binaries with a small q value.
Finally, \citet{Thompson+2021} suggested to determine the masses of the components of unresolved binary systems with the main sequence components in the open clusters by comparing the observed magnitudes in different photometric bands (including ones in the infrared range) with synthetic spectral energy distributions of stars.\\
Nevertheless, multi-filter pseudo colour photometric diagrams, in our opinion, allows to search for  unresolved binary systems more effectively, and  from this to obtain the characteristics of the distribution of binary stars in clusters.

In fact, in our previous study \citep{Malofeeva+2022}, we introduced a novel method to evaluate the fraction of unresolved binary stars and  higher multiplicity systems which exploits the pseudo-color diagram (H-W2)-W1 vs W2-(BP-K).
This pseudo-color diagram allows us to investigate the population of binaries and multiples in the relatively narrow primary component mass range around the solar mass.
This is because for higher masses, single stars and  multiple systems do not separate in this diagram.
Besides, for lower mass range the theoretical isochrones of \citet{Bressan+2012} do not reproduce the cluster main sequence satisfactorily.

The Pleiades star cluster was used in \citet{Malofeeva+2022} as a test-bench of the method.
We investigated the Pleiades star cluster in the primary component mass range between 0.5 and 1.8 $M_{\odot}$.
The binary star ratio was found to lie between 0.54$\pm$0.11 and 0.70$\pm$0.14.
The ratio of systems with a multiplicity larger than two was found to be between 0.10$\pm$0.00 and 0.14$\pm$0.01.
The distribution of the component mass ratio q was fitted by a power law with the exponent between -0.53$\pm$0.10 and -0.63$\pm$0.22.
Finally, \citet{Malofeeva+2022} stressed that below 0.5 $M_{\odot}$ a large number of brown dwarfs among secondary components is expected in the Pleiades.

However, we realized afterwards that the fitting of the $q$ distribution for Pleiades by a power law in \citet{Malofeeva+2022} could be significantly improved.
In fact, in that analysis, data on the numbers of stars were distributed using variable width $q$ data, and the difference in the bin size was not taken into account explicitly.
Therefore, we decided to revisit the fitting of the Pleiades data-set using a more robust approach.

First of all, in the present study we use automatic star counts for the various $q$ bins.
Secondly, we apply a bootstrapping method to evaluate the uncertainty of star numbers in the different $q$ bins.
And, finally, we model the distribution of the triple and quadruple systems in the (H-W2)-W1 vs W2-(BP-K) diagram in order to better constrain the region occupied by them and having the primary component in the desired mass range.

As a consequence, the layout of the paper is as follows.
Section~2 is devoted to the revision of the Pleiades results.
In Section~3 we describe the new findings for the Alpha Persei and Praesepe star clusters.
Section~4, then, is dedicated to the investigation of the star cluster NGC 1039.
Finally, Section~5 summarizes our results.

\section{A new look at the Pleiades star cluster} \label{sec:Pleiades}

Following \citep{Malofeeva+2022}, we used the diagram (H-W2)-W1 vs W2-(BP-K) to search for unresolved binary and multiple stars among the Pleaides probable cluster members.
We remind the reader that the sample was constructed as the intersection of two data-sets.
The first one comes from \citet{Danilov&Seleznev2020} and 
it includes stars inside a circle of 10.9 degrees wide up to $G=18$
magnitude.
The probability of being a cluster member is 95\% or higher, and the completeness is around 90\%.
The second source comes from \citet{Lodieu+2019}, and from this
we took the photometric data.

Fig.~1 shows the diagram for the selected stars with superimposed theoretical lines of constant $q$ values.
The right-most line corresponds to $q=0$, {\it i.e.} the
single stars sequence.
The left-most line corresponds to $q=1$, and it is the equal component mass binary sequence. 
We plotted these theoretical lines using isochrone tables from \citet{Bressan+2012} and following the procedure described in \citet{Malofeeva+2022}.
The cluster parameters $\log{t} = 8.116$, metallicity $[M/H]=0.032$, extinction $A_V = 0.168$, and the distance to the cluster of 135 pc are extracted from \citet{Dias+2021}.

\begin{figure}
    \centering
    \includegraphics[width=14cm]{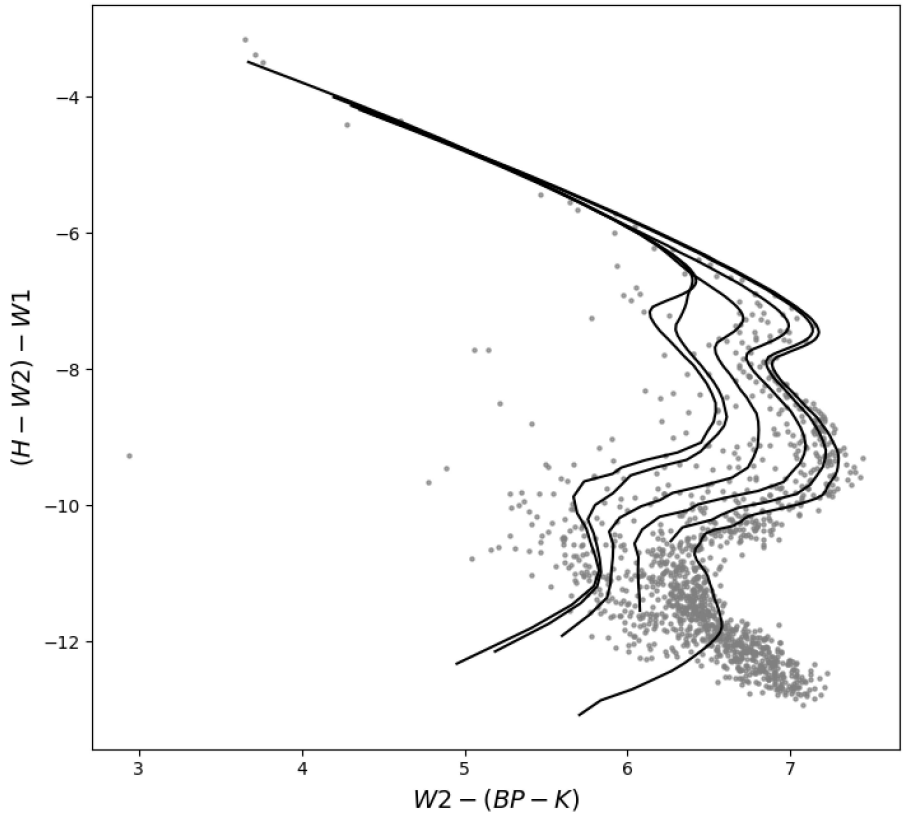}
    \caption{W2-(BP-K) vs (H-W2)-W1 diagram for the sample of the Pleiades probable members. Solid lines are for constant $q$ values. These lines of constant $q$ values refer to \citet{Bressan+2012} isochrone with $\log{t}=8.116$. $q$ values are: 0.0, 0.2, 0.4, 0.6, 0.8, 1.0 (from the right to the left).  }
     \label{Pleiades_diagram}
\end{figure}

A quick glance at Fig.~1 shows that the lines of constant $q$ values overlap in the upper part of the diagram.
On the other hand, in the lower part of the diagram the lines of constant $q$ values do not coincide well with the cluster sequence, as already stressed in \citet{Malofeeva+2022}.
Finally, one can notice that the various {bf isolines} terminates sharply in the lower part of the diagram due to the lack of data on stars with masses lower than 0.1 $M_{\odot}$ in the isochrone tables of \citet{Bressan+2012}.

Therefore, when counting stars in this diagram of Fig.~1, we are limited by the mass of the primary component in the range from 0.5 to 1.8 $M_{\odot}$.

We introduced some improvements with respect to the original procedure developed in \citet{Malofeeva+2022}.
Firstly, we determined more accurately the upper and lower boundaries of the region of stars with a multiplicity greater than two (dashed lines in Fig.~2 leftwards of the $q=1$ line).
To achieve this, we have modelled 600 triple and 600 quadruple star systems with the masses of the primary component from 0.5 $M_{\odot}$ to 1.8 $M_{\odot}$, and the masses of other components from 0.1 $M_{\odot}$ up to the mass of the main component.
The masses of the primary components were distributed as a geometrical progression (with the use of {\it numpy.geomspace} routine).
This distribution is close to a uniform one in log space, but all the mass values are unique (not repeated), and the number of the low-mass stars is larger than in the uniform case.
Note that the main goal in this case is to show in the best possible way how binary and higher multiplicity systems are distributed.
Realistic mass distributions are ill-suited for this, since they give a much smaller number of massive stars compared to low-mass ones.
The goal of this choice is to achieve a better representation of the region occupied by the triple and quadruple systems with the primary component mass in the range of interest.
Fig.~2 shows the location of the unresolved triple and quadruple stars in the diagram.
The blue and orange lines correspond to triple and quadruple unresolved systems with equal components.
The thin black lines are lines of equal $q$ values in the same order as in Fig.~1.
It can be noticed that the regions of triple and quadruple systems overlap with the region of binary stars.
Therefore, some of the Pleiades stars identified as binary may actually turn out to be triple or quadruple stars.
A a consequence, the estimate of the binary star number would be an upper one.
In turn, the estimates of the number of triple and quadruple stars would be the lower ones, if we count stars between the left thin black line and blue line for triple stars, and between blue line and orange line for quadruple stars.

\begin{figure}
    \centering
    \includegraphics[width=14cm]{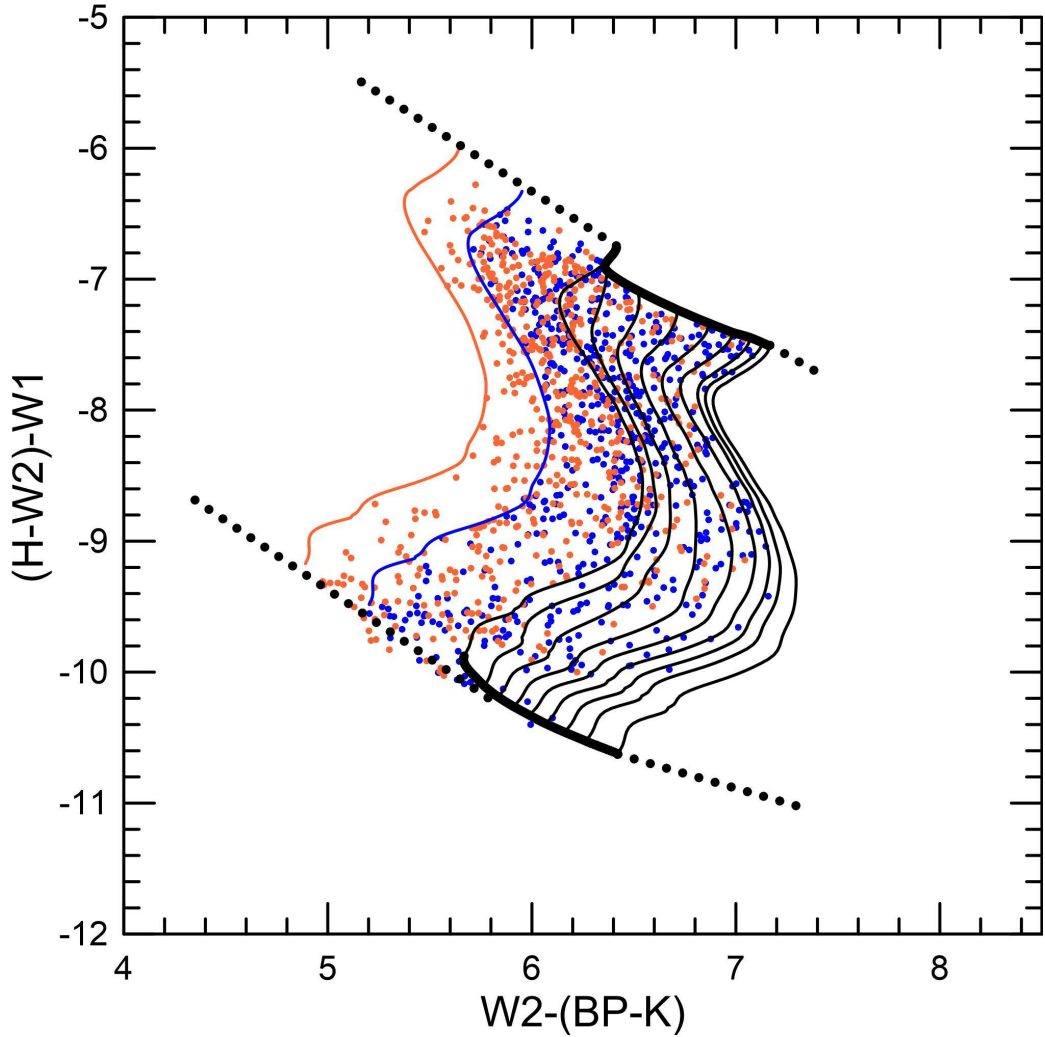}
    \caption{(H-W2)-W1 vs W2-(BP-K) diagram for triple and quadruple unresolved stars with the primary component in the range between 0.5 $M_{\odot}$ and 1.8 $M_{\odot}$.
    Blue dots are triple stars.
    Orange dots are quadruple stars.
    The blue and orange lines correspond to triple and quadruple unresolved systems with equal components.
    The thin black lines are lines of equal $q$ values in the following order (from the right to the left): $q=0,0.2,0.3,0.4,0.5,0.6,0.7,0.8,1.0$.
    }
     \label{triples}
\end{figure}

Fig.~3 shows the pseudo-color diagram with the upper and lower boundaries of the star counting area.
The upper line corresponds to binary and multiple stars with the primary component of 1.8 $M_{\odot}$. 
The lower line corresponds to binary and multiple stars with the primary component of 0.5 $M_{\odot}$.

\begin{figure}
    \centering
    \includegraphics[width=14cm]{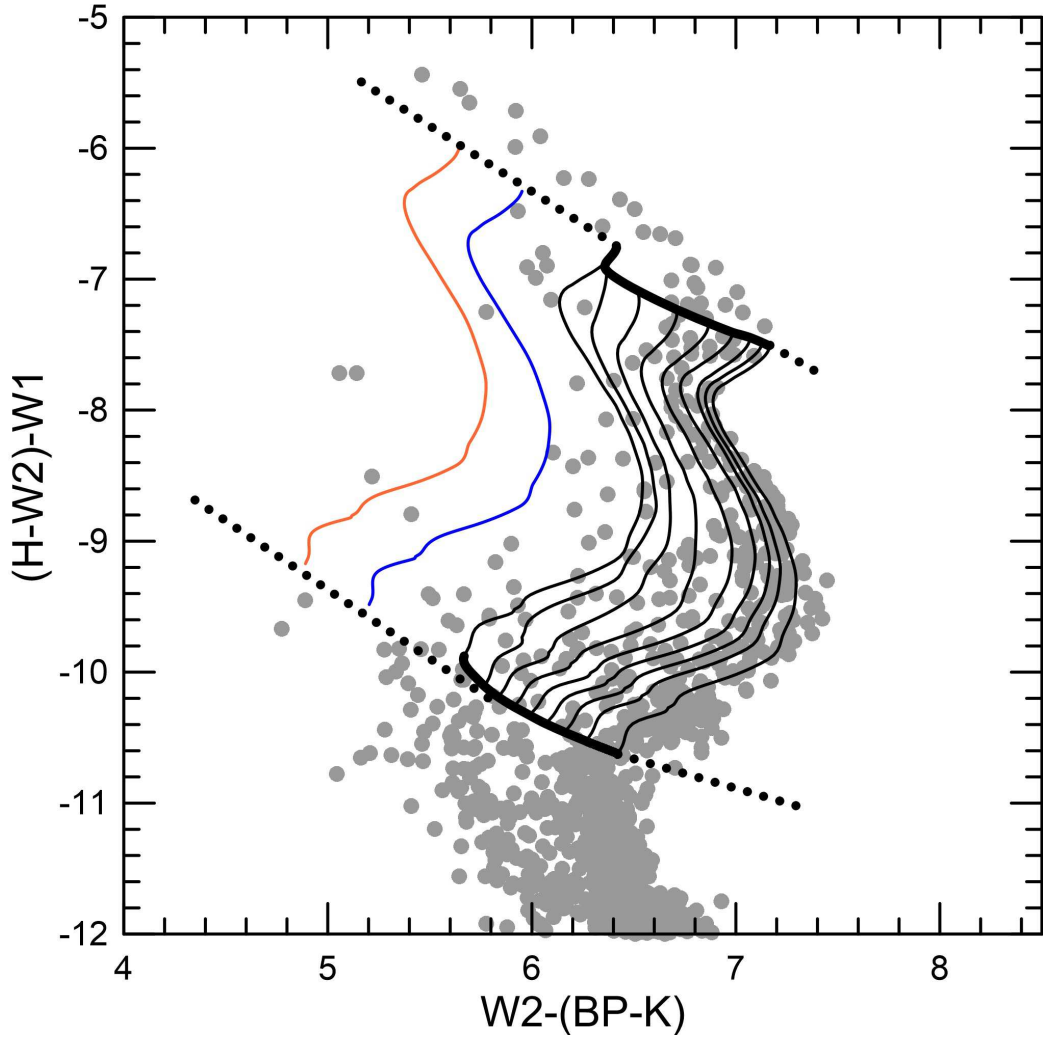}
    \caption{(H-W2)-W1 vs W2-(BP-K) diagram for Pleaides probable members (gray dots) with superimposed lines of constant $q$ values (as in Fig.2), lines for multiple systems with equal components (as in Fig.2), and with the boundaries (thick solid and dashed lines) of the area of star counts.}
     \label{Pleiades_count}
\end{figure}

The second improvement is that we have made the bins covering equal intervals in $q$ for a correct construction of the $q$-distribution $f(q)$. 
It turns out that the $f(q)$ has a maximum at $q\sim 0.3-0.4$ (see below).

The counting of the single stars was performed in the same way as in the previous work \citep{Malofeeva+2022}.
Since single stars are plotted on the diagram with errors, they fall on both sides of the theoretical line of single stars ($q=0$). Therefore, we propose to consider them in two ways.
In the first case (case a), and with reference to Fig.~3, only points rightwards of the single stars ($q=0$) sequence are considered to be single stars.
In the second case (case b), we add points with $q$ from 0 to 0.2 to the single star sample.
The upper and lower boundaries of the region of single stars are drawn as tangents to the lines of binary stars with the primary components with the mass of 0.5 $M_{\odot}$ and of 1.8 $M_{\odot}$.
Star counts were performed automatically.
For every $q$ bin, the position of each star relative to the lines bounding this bin was determined by linear interpolation.

Another important improvement with respect to our previous study \citep{Malofeeva+2022} is the use of a bootstrap method to account for photometric errors.
In detail, the observational data points were re-plotted randomly according to a normal distribution with the mean and standard deviation taken from the catalog for any given star, and the procedure of the star counts repeated.
Bootstrapping was repeated 100 times, then the average values of the number of stars in the $q$ bins and their standard deviations were computed.
The results of star counts in different $q$ bins are shown in Tab.~1 for the Pleiades, Alpha Persei, and Praesepe star clusters.

\begin{deluxetable*}{cccc}
\tablenum{1}
\tablecaption{The results of counts of stars with different values of $q$ in Pleiades, Alpha Per, and Praesepe clusters\label{tab:counts}}
\tablewidth{0pt}
\tablehead{
\colhead{The range of $q$} & \colhead{Number of stars} & \colhead{Number of stars} & \colhead{Number of stars} \\
\colhead{} & \colhead{Pleiades} & \colhead{Alpha Per} & \colhead{Praesepe} \\}
\startdata
to the right of $q=0$ & 112$\pm$5  & 123$\pm$6    &  89$\pm$4    \\
$0<q<0.2$             & 74$\pm$6   &  60$\pm$6    &  35$\pm$5    \\
$0.2<q<0.4$           & 81$\pm$5   &  65$\pm$5    &  47$\pm$5    \\
$0.4<q<0.6$           & 69$\pm$5   &  54$\pm$4    &  19$\pm$2    \\
$0.6<q<0.8$           & 34$\pm$3   &  24$\pm$2    &  15$\pm$2    \\
$0.8<q<1.0$           & 10$\pm$2   &   6$\pm$2    &   6$\pm$2    \\
triple systems        & 29$\pm$2   &  20$\pm$2    &  13$\pm$2    \\
quadruple systems     &  3$\pm$1   &   2$\pm$1    &   1$\pm$1    \\
\enddata
\end{deluxetable*}

Fig.~1 indicates also a systematic deviation of the star sequence rightwards the $q=0$ line, and below $(H-W_2)-W_1 \sim -8.4$  (exactly as in our previous work \citet{Malofeeva+2022}).
We envisage two possible reasons for this occurrence.
Firstly of all this is probably the cumulative effect of increasing the number of stars and the photometric errors (see Fig.3a from \citet{Malofeeva+2022}).
Beside, it might also originate from the isochrone fitting procedure, which is effective in the upper part of the diagrams, but exhibits a clear mis-match with the cluster sequence downwards $-8.4$.
We anticipate that this occurs in all the clusters under consideration (see below).

This deviation can artificially increase single star counts.
However, this is partially accounted for by our procedure because the typical values of the pseudo-color' errors are comparable with the star count variations.\\

\noindent
Fig.~4 shows the distribution function $f(q)$ for the first case (single stars are only to the right of the line $q=0$, the solid line, case (a)) and the second case (stars in the bin $q\in[0;0.2]$ are added to single stars,  the dashed line, case (b)).
Since there is a clear maximum around $q\sim 0.3-0.4$, we followed  \citet{Kouwenhoven+2009} and performed the fitting by a Gaussian function using a non-linear least squares method with errors (using the {\it curve\_fit} routine from the Python package {\it scipy.optimize}).

\begin{figure}
    \centering
    \includegraphics[width=14cm]{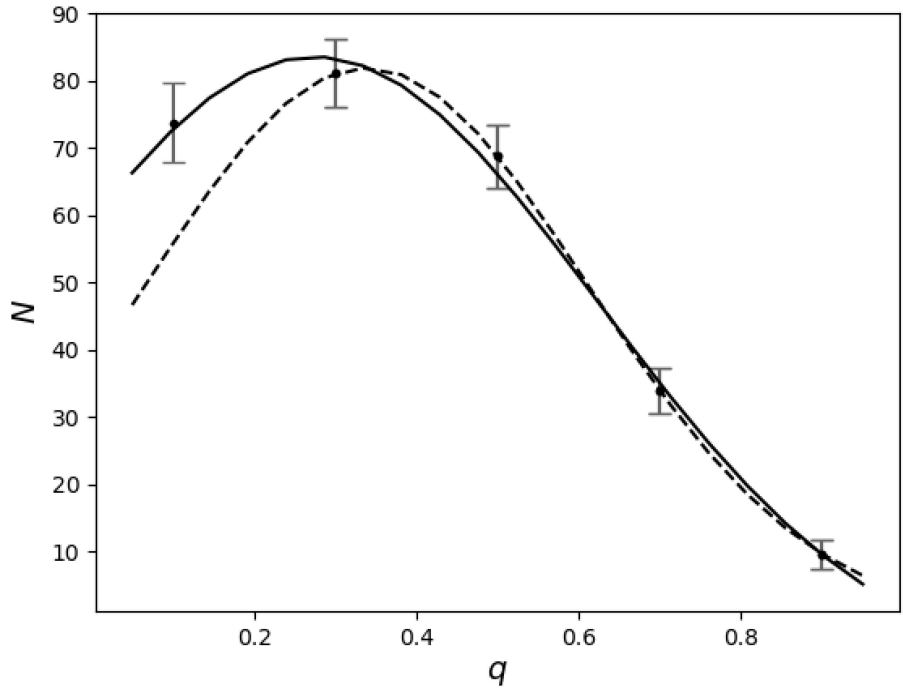}
    \caption{The distribution of the component mass ratio $q$ and its fitting by a Gaussian for the Pleiades.
    Solid line (case (a)) --- stars with $0<q<0.2$ are considered as binaries; dashed line (case (b))--- stars with $0<q<0.2$ are added to singles.
    }
     \label{q_Pleiades}
\end{figure}

\begin{equation}
\label{Gauss}
dN/dq\sim exp[-(q-\mu_q)^2/2\sigma_q^2]
\end{equation}

\noindent The mean values and standard deviations were equal to $\mu=0.27\pm0.03$, $\sigma=0.35\pm0.07$ for case (a) and $\mu=0.34\pm0.06$, $\sigma=0.27\pm0.07$ for case (b).
For the parameter $\alpha$

\begin{equation}
\label{alpha}
\alpha=\frac{N_{binaries}+N_{triples}+N_{quadruples}}{N_{singles}+N_{binaries}+N_{triples}+N_{quadruples}}\; ,
\end{equation}

\noindent where $\beta$ is:

\begin{equation}
\label{beta}
\beta=\frac{N_{triples}}{N_{binaries}+N_{triples}+N_{quadruples}}\; ,
\end{equation}

\noindent and $\gamma$ is: 

\begin{equation}
\label{gamma}
\gamma=\frac{N_{quadruples}}{N_{binaries}+N_{triples}+N_{quadruples}}\; ,
\end{equation}

\noindent 
we obtained the values listed in Tab.~3 along with the parameters of $q$ distribution (see below).
This result is in line with our previous study \citep{Malofeeva+2022}.

The conclusion of \citet{Malofeeva+2022} on the presence of a large number of the binary systems with the very-low-mass secondary components (quite probably, brown dwarfs) in the Pleiades remains unaltered.
We discuss this point in more details below.
 
\section{Binary and multiple stars population in Alpha Per and Praesepe star clusters} \label{Alpha Per and Praesepe}

We obtained the parameters of the population of binary and multiple stars in the Alpha Persei and Praesepe clusters in the same way as in the Pleiades, exploiting the photometric diagram (H-W2)-W1 – W2-(BP-K).
For the Alpha Persei cluster, we used an isochrone with $\log{t}=7.921$, metallicity $[M/H]=0.158$, extinction $A_V=0.324$, and the distance to the cluster $r=177$ pc.
For the Praesepe cluster, we used an isochrone  with $\log{t} = 8.882$, metallicity $[M/H]=0.196$, extinction $A_V = 0.032$, and the distance to the cluster $r=186$ pc.
These sets of cluster parameters were extracted from \citet{Dias+2021}.

Samples of probable cluster members of the Alpha Persei and Praesepe clusters were built from the catalog of \citet{Lodieu+2019}.
We excluded stars with $G>18$ mag.
Some stars below the cluster main sequence on the CMD also were excluded.
The sample of the probable cluster members of the Alpha Persei cluster is the intersection of the \citet{Lodieu+2019} sample and the cluster sample {\it C2}, obtained in \citet{Nikiforova+2020} with {\it DBSCAN} to avoid the contamination by stars belonging to the stellar stream in the vicinity of the Alpha Persei cluster.
Eventually, the sample of the Alpha Persei probable members contains 956 stars, whereas the sample of the Praesepe probable members contains 2200 stars.
The theoretical sequences of single and binary stars, namely the lines of  constant $q$, the upper and lower boundaries of the regions for star counts, and the lines for multiple systems with equal components, were plotted in the same way as for the Pleiades.
Figs.~5 and ~6 contain the resulting color-color diagrams with theoretical lines and member stars of the Alpha Persei and the Preasepe cluster, respectively. 
The lower and upper bounds for Alpha Persei correspond to stars with masses of the main components of 0.5 $M_{\odot}$ and 1.9 $M_{\odot}$, while for the Praesepe --- of 0.7 $M_{\odot}$ and 1.4 $M_{\odot}$.

\begin{figure}
    \centering
    \includegraphics[width=14cm]{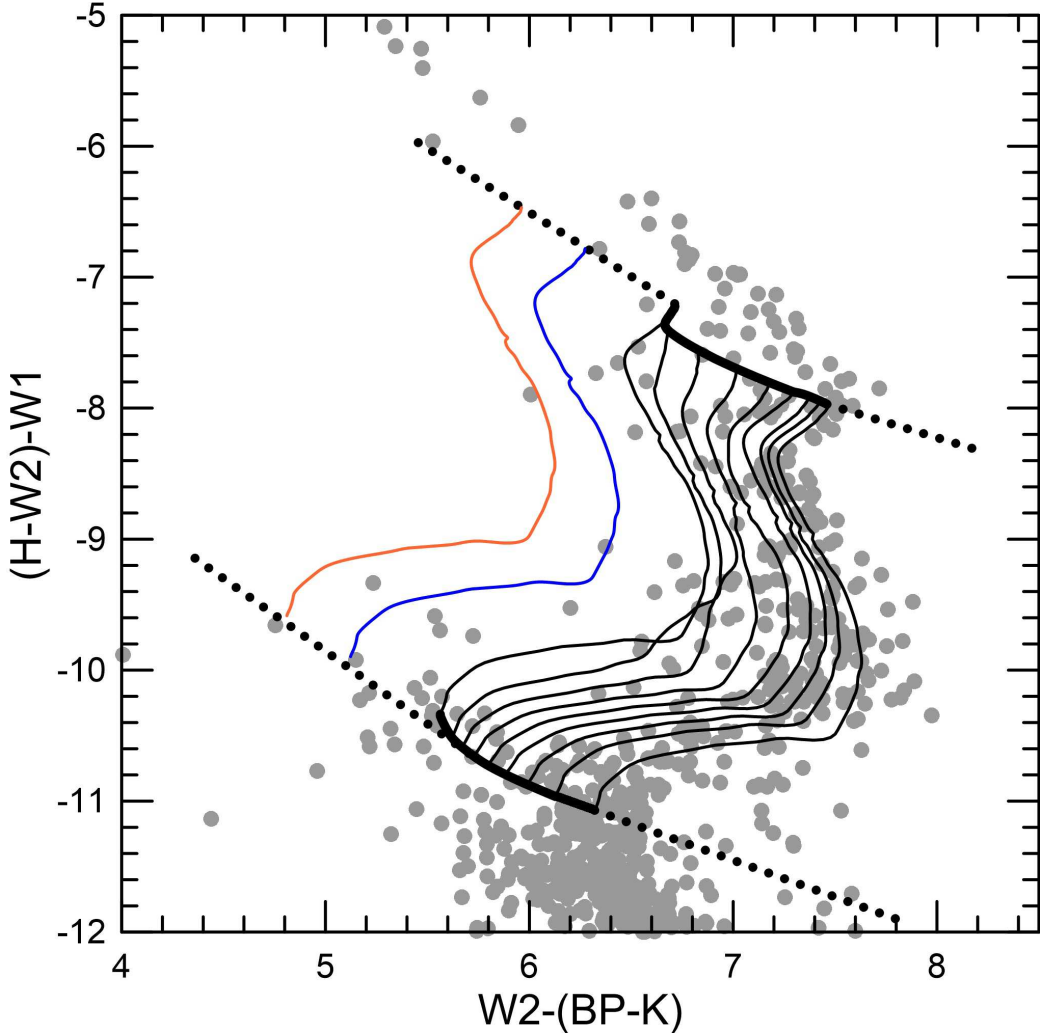}
    \caption{The (H-W2)-W1 vs W2-(BP-K) diagram for the probable members of the Alpha Persei (gray dots).
    Lines and designations are the same as in Fig.~3.
    The upper lines corresponds to the binary and multiple stars with the primary component of 1.9 $M_{\odot}$.
    The lower lines corresponds to the binary and multiple stars with the primary component of 0.5 $M_{\odot}$. }
     \label{Alpha_Persei_count}
\end{figure}

\begin{figure}
    \centering
    \includegraphics[width=14cm]{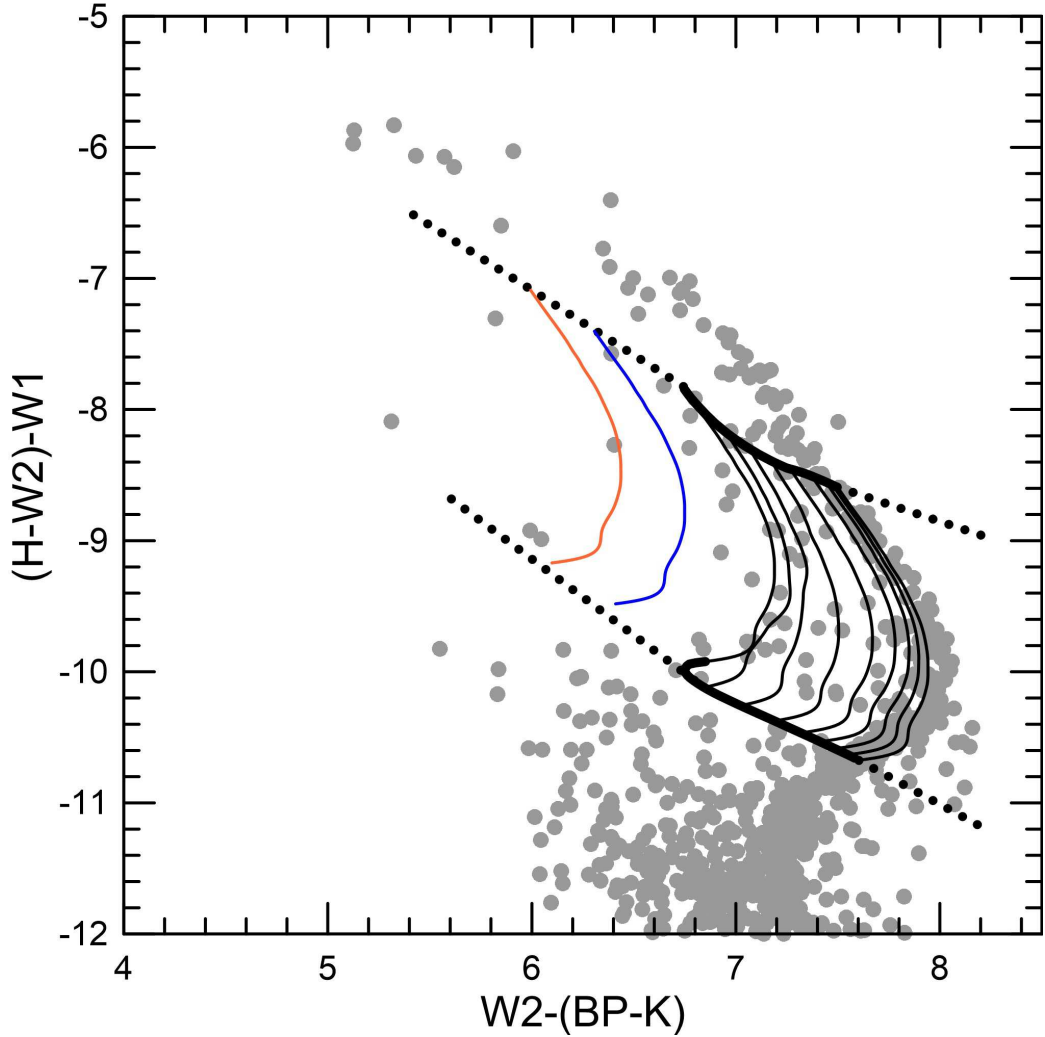}
    \caption{The (H-W2)-W1 vs W2-(BP-K) diagram for the probable members of the Praesepe (gray dots).
    Lines and designations are the same as in Fig.~3.
    The upper lines corresponds to the binary and multiple stars with the primary component of 1.4 $M_{\odot}$.
    The lower lines corresponds to the binary and multiple stars with the primary component of 0.7 $M_{\odot}$. }
     \label{Praesepe_count}
\end{figure}

The procedure of star counts in the $q$ bins and multiple bins, and the accounting of photometry errors for the Alpha Persei and Praesepe clusters were exactly the same as for the Pleiades (see above).
The results of star counts are then listed in Tab.~1.
On the other hand, the distribution functions of the $q$ parameter for Alpha Persei is shown in Fig~7.
For the Praesepe cluster we faced a problem since excluding the point corresponding to the $q$ bin $[0;0.2]$ the fitting routine could not return acceptable parameters for the Gaussian curve.
To circumvent this problem we performed star counts for Praesepe cluster in  narrower $q$ bins (of 0.1 width) to increase the number of points for the fitting.
The results of star counts for Praesepe with thin bin choice are listed in Tab.~2.
Then, to provide the equal bins of 0.2 for fitting, in line with the other clusters, we used the sum of two adjacent narrower bins.

\begin{deluxetable*}{cc}
\tablenum{2}
\tablecaption{The results of counts of stars in the narrow $q$ bins in the Praesepe cluster\label{tab:Praesepe_counts}}
\tablewidth{0pt}
\tablehead{
\colhead{The range of $q$} & \colhead{Number of stars}  \\
\colhead{} &  \colhead{Praesepe} \\}
\startdata
$0.2<q<0.3$          &  27$\pm$4    \\
$0.3<q<0.4$          &  20$\pm$3    \\
$0.4<q<0.5$          &  11$\pm$2    \\
$0.5<q<0.6$          &   8$\pm$2    \\
$0.6<q<0.7$          &   8$\pm$1    \\
$0.7<q<0.8$          &   7$\pm$2    \\
\enddata
\end{deluxetable*}

For Praesepe, moreover, the (b) case, when we consider stars in the bin $q\in[0;0.2]$ as single stars, corresponds to assuming that the number of the binary stars in that bin is equal to zero.
This way we could solve the problem for the Praesepe cluster (see Fig.~8 for the distribution of $q$ and its fitting).
Tab.~3 lists the parameters of distribution of $q$ (fitted by Gaussian Eq.\ref{Gauss}) and the values of $\alpha$, $\beta$, and $\gamma$ for clusters.

\begin{figure}
    \centering
    \includegraphics[width=14cm]{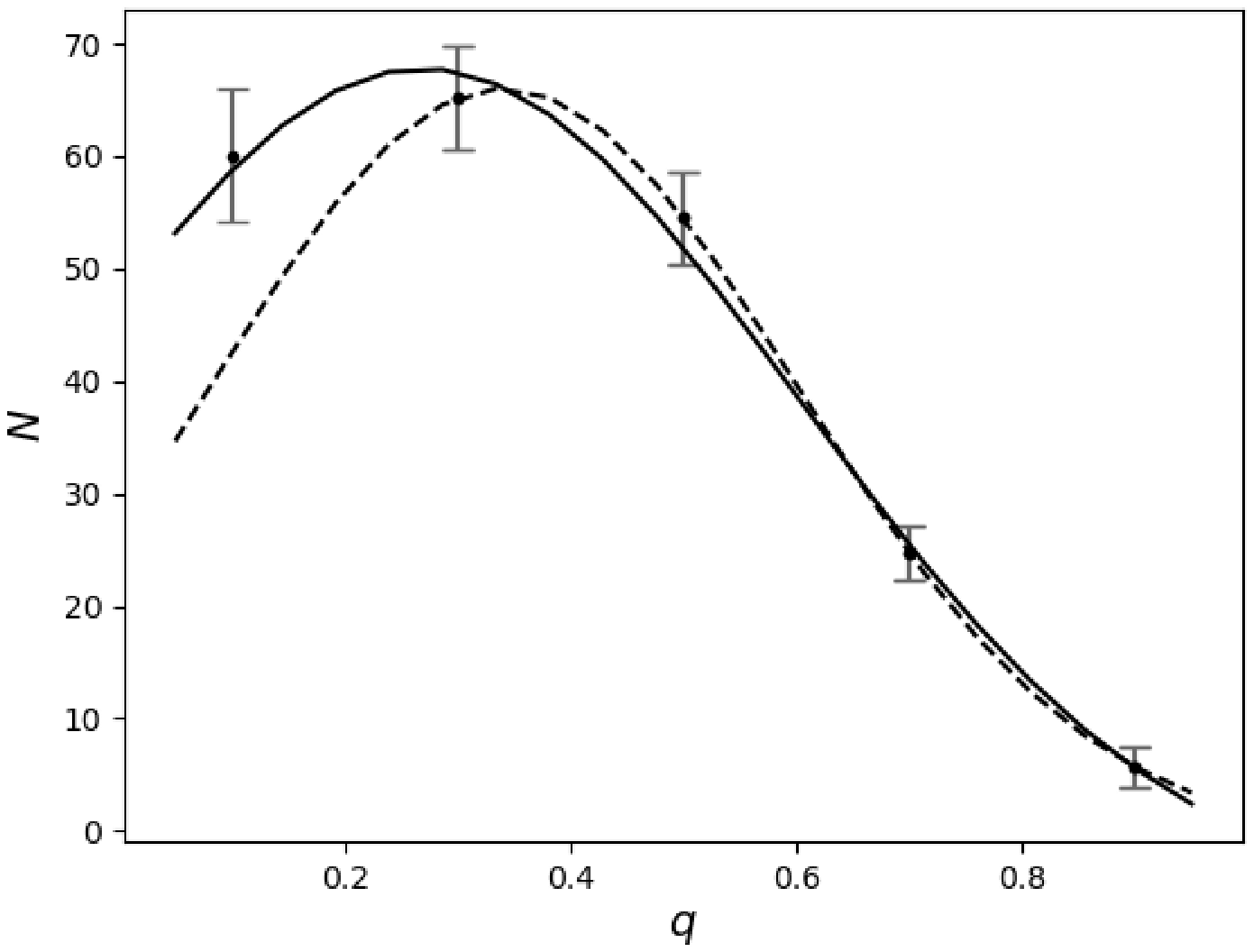}
    \caption{The distribution of the component mass ratio $q$ and the Gaussian fit for the Alpha Persei cluster.
    Designations are the same as in Fig.~4.
    }
     \label{q_Alpha_Persei}
\end{figure}

\begin{figure}
    \centering
    \includegraphics[width=14cm]{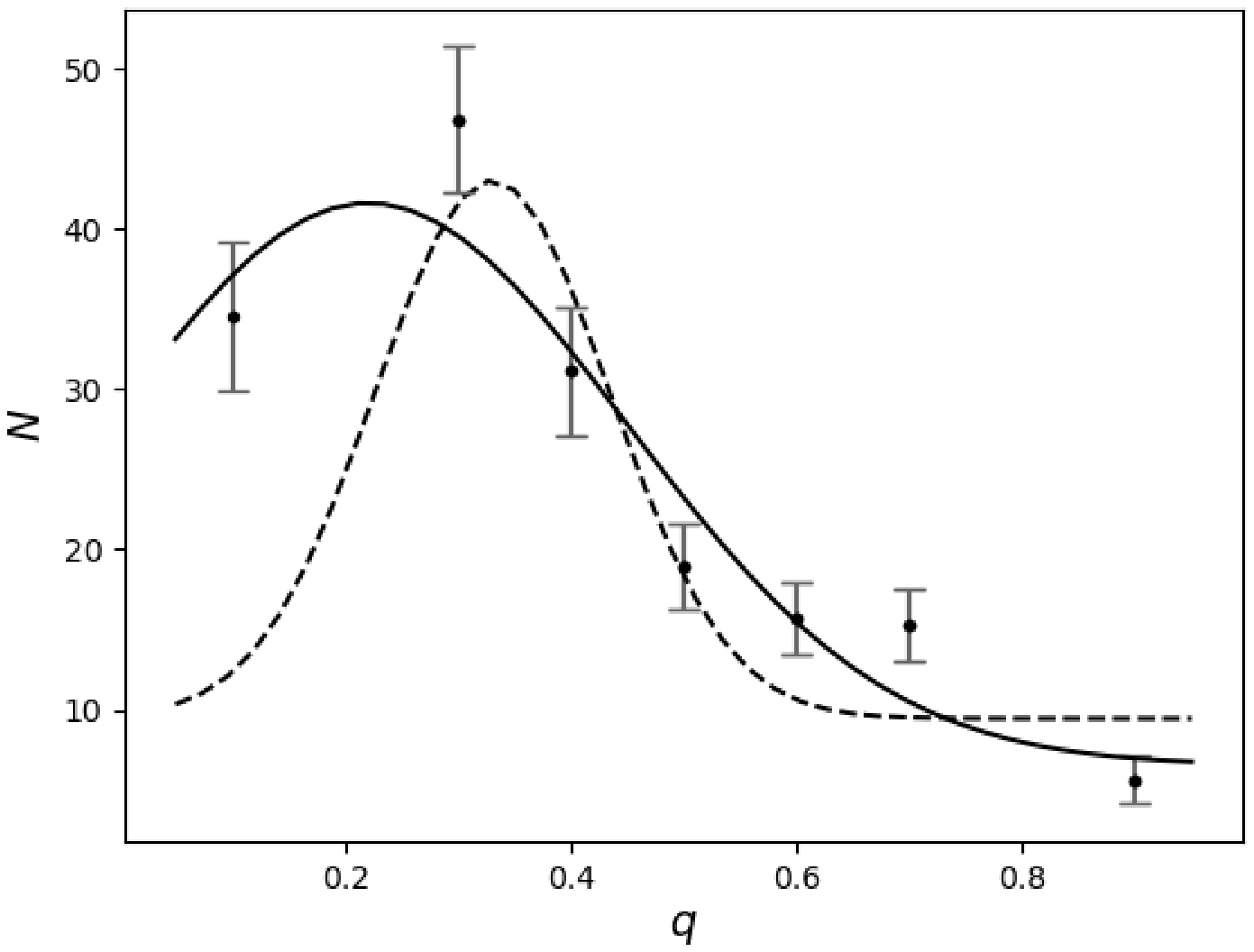}
    \caption{The distribution of the component mass ratio $q$ and the Gaussian fit for the Praesepe cluster.
    Designations are the same as in Fig.~4.
    }
     \label{q_Praesepe}
\end{figure}

\begin{deluxetable*}{ccccc}
\tablenum{3}
\tablecaption{Parameters of the binary and multiple star population in the clusters\label{tab:q_distr}}
\tablewidth{0pt}
\tablehead{
\colhead{Parameter} & \colhead{Pleiades} & \colhead{Alpha Persei} & \colhead{Praesepe} & \colhead{NGC 1039}  \\
}
\startdata
$\mu$ (1 case)    & 0.27$\pm$0.03 & 0.27$\pm$0.03 & 0.22$\pm$0.04 & 0.51$\pm$0.02     \\
$\mu$ (2 case)    & 0.34$\pm$0.06 & 0.34$\pm$0.05 & 0.33$\pm$0.02 & 0.52$\pm$0.01     \\
$\sigma$ (1 case) & 0.35$\pm$0.07 & 0.32$\pm$0.05 & 0.23$\pm$0.04 & 0.23$\pm$0.02     \\
$\sigma$ (2 case) & 0.27$\pm$0.07 & 0.25$\pm$0.05 & 0.10$\pm$0.02 & 0.20$\pm$0.01     \\
$\alpha$ (1 case) & 0.73$\pm$0.03 & 0.65$\pm$0.03 & 0.60$\pm$0.04 & 0.70$\pm$0.02     \\
$\alpha$ (2 case) & 0.55$\pm$0.02 & 0.48$\pm$0.02 & 0.45$\pm$0.03 & 0.63$\pm$0.02     \\
$\beta$ (1 case)  & 0.10$\pm$0.01 & 0.09$\pm$0.01 & 0.10$\pm$0.02 & 0.11$\pm$0.02     \\
$\beta$ (2 case)  & 0.13$\pm$0.01 & 0.12$\pm$0.01 & 0.13$\pm$0.02 & 0.13$\pm$0.02     \\
$\gamma$ (1 case) &0.010$\pm$0.003&0.009$\pm$0.004&0.007$\pm$0.007& 0.02$\pm$0.01     \\
$\gamma$ (2 case) &0.013$\pm$0.004&0.012$\pm$0.006&0.010$\pm$0.010& 0.02$\pm$0.01     \\
\enddata
\end{deluxetable*}

\section{Binary and multiple stars population in NGC 1039} \label{NGC 1039}

We have searched for unresolved binary and multiple stars in the open star cluster NGC 1039 applying the procedure described above to a sample of probable cluster members.
This sample was compiled starting from the catalog of \citet{Cantat-Gaudin+2020} with the photometry from Gaia DR2 on condition $G\leqslant18$.
We have added the necessary photometric bands from the catalogs of 2MASS \citep{2MASS} and WISE \citep{WISE} surveys.
We have excluded stars with the membership probability less than 50\%.
As a result, the final sample contains 553 stars.

\begin{figure}
    \centering
    \includegraphics[width=13 cm, keepaspectratio=1]{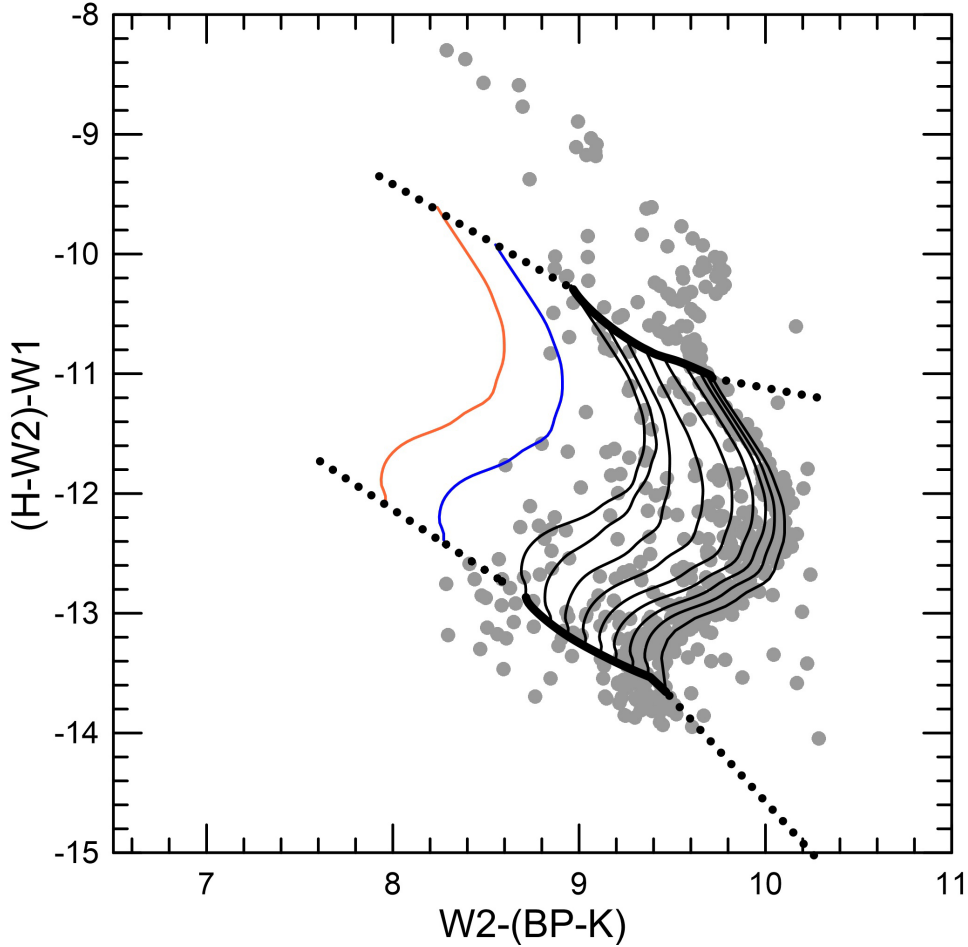}
    \caption{The diagram (H-W2)-W1 vs W2-(BP-K) for the probable members of NGC 1039 (gray dots) with the thin black lines for constant q values of 0.0, 0.2, 0.3, 0.4, 0.5, 0.6, 0.7, 0.8, 1.0 (from the right to left), and with the boundaries of the area of star counts.
    An upper line corresponds to binary and multiple stars with the primary component of 1.3 $M_{\odot}$.
    A lower line corresponds to binary and multiple stars with the primary component of 0.5 $M_{\odot}$.
    The blue and orange lines designate triple and quadruple systems with equal components, respectively. }
    \label{ngc1039_cid}
\end{figure}

We construct the theoretical sequence of single stars and lines of the constant $q$ values using the isochrone tables of \citet{Bressan+2012} with the age logarithm $\log{t}=8.116$, metallicity $[Fe/H]=-0.006$, and extinction $A_V=0.328$.
These cluster parameters and the distance to the cluster of $r=507$ pc were again extracted from \citet{Dias+2021}.
Fig.\ref{ngc1039_cid} shows the diagram with the theoretical lines, the cluster stars and the lower and upper limits corresponding to systems with the primary component mass of 0.5 $M_{\odot}$ and 1.3 $M_{\odot}$, respectively.
As for the case of the Pleiades (see above), we have modelled the distribution of the triple and quadruple systems with the primary component mass in the same range in order to better outline the region occupied by multiple systems.

\begin{figure}
    \centering
    \includegraphics[width=13 cm, keepaspectratio=1]{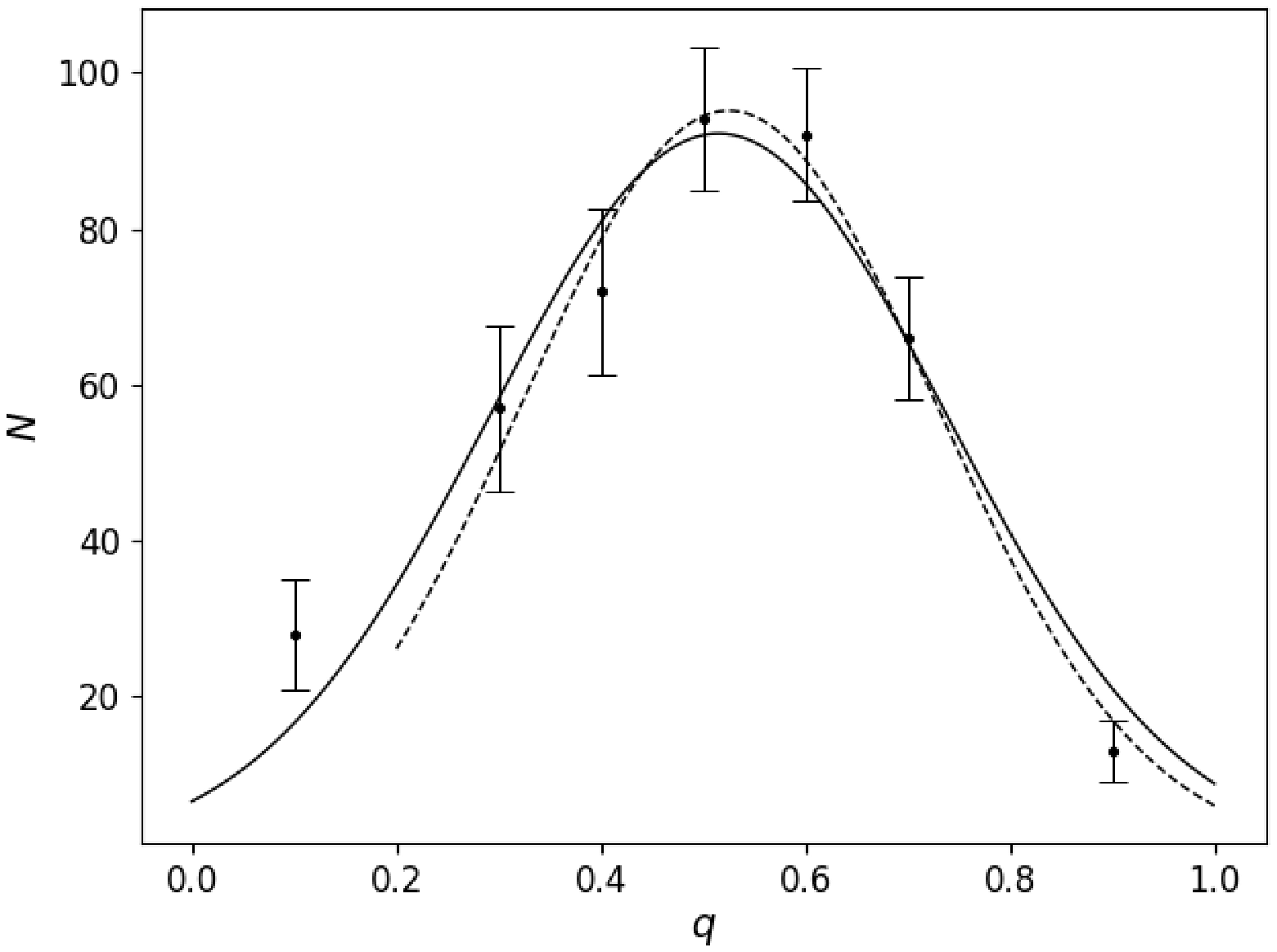}
    \caption{The distribution of the component mass ratio $q$ and its fitting by a Gaussian for NGC 1039.
    Solid line is for case (a) while  dashed line is for case (b).}
    \label{ngc1039_d}
\end{figure}

The procedure of counting  multiple, binary and single stars and the accounting for photometric errors are the same as described above for the Pleiades, Alpha Persei, and Praesepe.
The only difference is that the boot-strapping procedure has been performed 300 times in this case, because we noted that 100 times were insufficient for this cluster to reach stable results.
The results of star counts are listed in Table \ref{Counts_ngc1039}, while
the distribution functions of the $q$ parameter are shown in Figure \ref{ngc1039_d}.
In order to keep the $q$ bins equal ($\Delta q=0.2$) the data from Table \ref{Counts_ngc1039} were stacked in the following way: $0.2<q<0.3$ and $0.3<q<0.4$, $0.3<q<0.4$ and $0.4<q<0.5$, $0.4<q<0.5$ and $0.5<q<0.6$ and so on. 
The distribution of $q$ was fitted with a Gaussian for two cases (including the point for the bin $0.0<q<0.2$ and without it).
The parameters of the Gaussian and the ratios of the binary and multiple stars are listed in the last column of Table~3.

\begin{deluxetable*}{cc}
\tablenum{4}
\tablecaption{Results of counts of stars with different values of $q$ and multiple stars in NGC 1039\label{Counts_ngc1039}}
\tablewidth{0pt}
\tablehead{
\colhead{The range of $q$} & \colhead{Number of stars NGC 1039} \\}
\startdata
to the right of $q=0$ & 109$\pm$9   \\
$0<q<0.2$             & 28$\pm$8   \\
$0.2<q<0.3$           & 27$\pm$7   \\
$0.3<q<0.4$           & 32$\pm$8    \\
$0.4<q<0.5$           & 42$\pm$8    \\
$0.5<q<0.6$           & 53$\pm$6    \\
$0.6<q<0.7$           & 40$\pm$6    \\
$0.7<q<0.8$           & 28$\pm$6    \\
$0.8<q<1.0$           & 13$\pm$4    \\
triple systems        & 30$\pm$6    \\
quadruple systems     &  5$\pm$3    \\
\enddata
\end{deluxetable*}

\section{Conclusions} \label{sec:Conclusions}

In this study, we investigated the population of binary and multiple stars in the Pleiades, Alpha Persei, Praesepe, and NGC 1039 Galactic star clusters.
To this purpose, we used a photometric diagram which employs two pseudo-colors constructed from stellar magnitudes in the pass-bands of visible and infrared wavelengths W2-(BP-K) vs (H-W2)-W1, originally proposed by \citet{Malofeeva+2022}.
The advantage of this color combination is that binary stars in this diagram pop clearly up and are well distinguished from single stars even for small values of the component mass ratio $q$.

The most important limitation of our investigation is that theoretical isochrones (the line of single stars, $q=0$) do not coincide well with the cluster main sequence for the whole stellar magnitude range.
Because of that, we were forced to limit the range of our exploration of this diagram to intermediate masses for the primary component:  $m_1\in[0.5;1.8]$ $M_{\odot}$ for Pleiades, $m_1\in[0.5;1.9]$ $M_{\odot}$ for Alpha Persei, $m_1\in[0.7;1.4]$ $M_{\odot}$ for Praesepe, and $m_1\in[0.5;1.3]$ $M_{\odot}$ for NGC 1039.

\begin{figure}
    \centering
    \includegraphics[width=19cm]{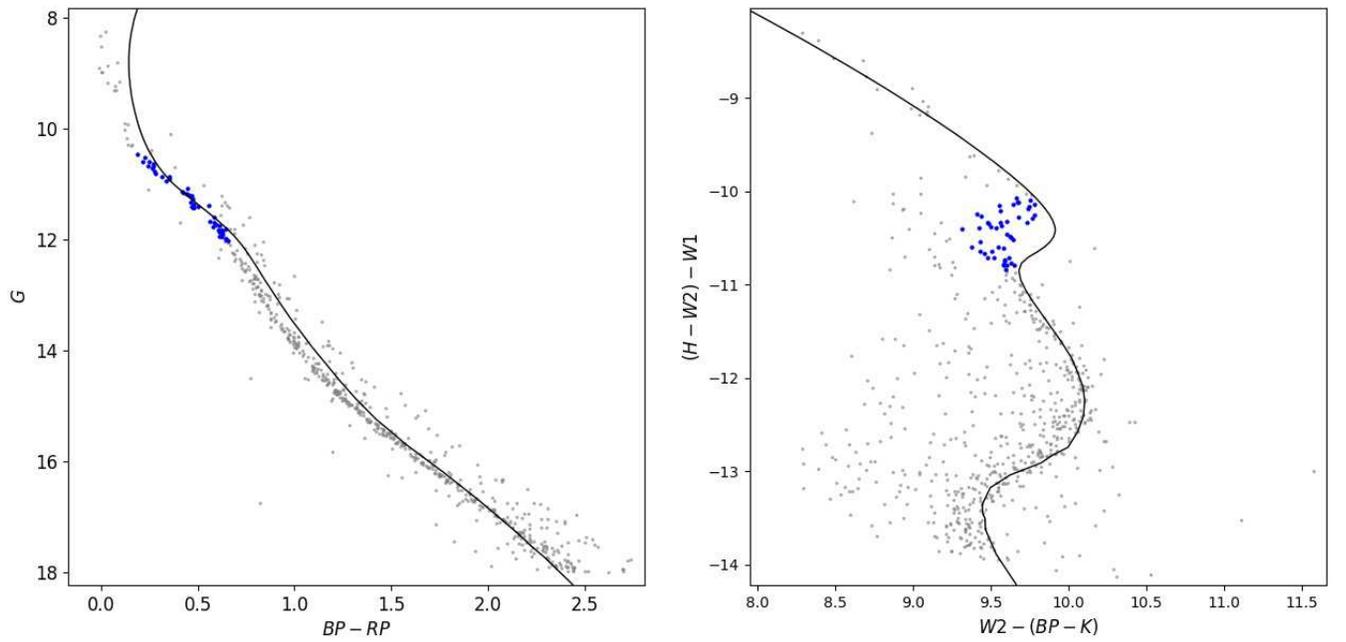}
    \caption{The NGC 1039 CMD (left), and the diagram W2-(BP-K) vs (H-W2)-W1 (right).
    The solid line represents the theoretical isochrone over-imposed for the set of cluster parameters from \citet{Dias+2021}.
    The blue dots mark the same stars on both diagrams.
    }
     \label{NGC 1039 CMD}
\end{figure}

We used the fundamental cluster parameters from the catalog of \citet{Dias+2021}.
As an illustration, Fig. \ref{NGC 1039 CMD} shows the CMD for NGC 1039 (the left panel) and the diagram W2-(BP-K) vs (H-W2)-W1 for this cluster (the right panel). 
In general the fitting is rather disappointing.
If one tries to adjust the upper part of the diagram, the lower part 
gets off, and vice versa.
The blue dots in Fig. \ref{NGC 1039 CMD} mark the same stars on both diagrams.
It is well visible that while on the CMD these stars are close enough to the isochrone, on the diagram W2-(BP-K) vs (H-W2)-W1 the same stars are far from the isochrone.
The reason is unclear, but it is because of this point that we had to limit the mass range of our investigation.

In the case of the Pleiades we improved on \citet{Malofeeva+2022} fitting of the $q$ distribution with the implementation of equal $q$ bins.
As a result, it turned out that a power law is not suitable because this distribution has a maximum for all clusters under investigation.
We used a Gaussian curve instead following  \citet{Kouwenhoven+2009}, with the parameters of the fitting listed in Table 3.
The mode of distribution ranges from 0.22 for Praesepe to 0.52 for NGC 1039.
The dispersion ranges between 0.10$\pm$0.02 for Praesepe and 0.35$\pm$0.07 for Pleiades.

The parameters of the $q$ distribution are close each except for the mode in NGC 1039 and the dispersion in Praesepe (for the case (b) single star number estimate; in this case the approximation includes the additional condition of the absence of binaries with $q\in[0;0.2]$).
The ratio of the binary and multiple stars and ratios of the triple and quadruple systems also are close for the sample clusters (see below).
We could not find any correlation of these parameters with the cluster age.

An important point is the relatively small number of binaries with equal mass components.
This seems to contradict the results of \citet{El-Badry+2019}.
A possible explanation for this contradiction is that we take into account the unresolved systems only.
Another possibility is that tight resolved binaries with similar components could have bad Gaia astrometric solutions and be missed in the samples of probable cluster members.\\

\begin{figure}
    \centering
    \includegraphics[width=19cm]{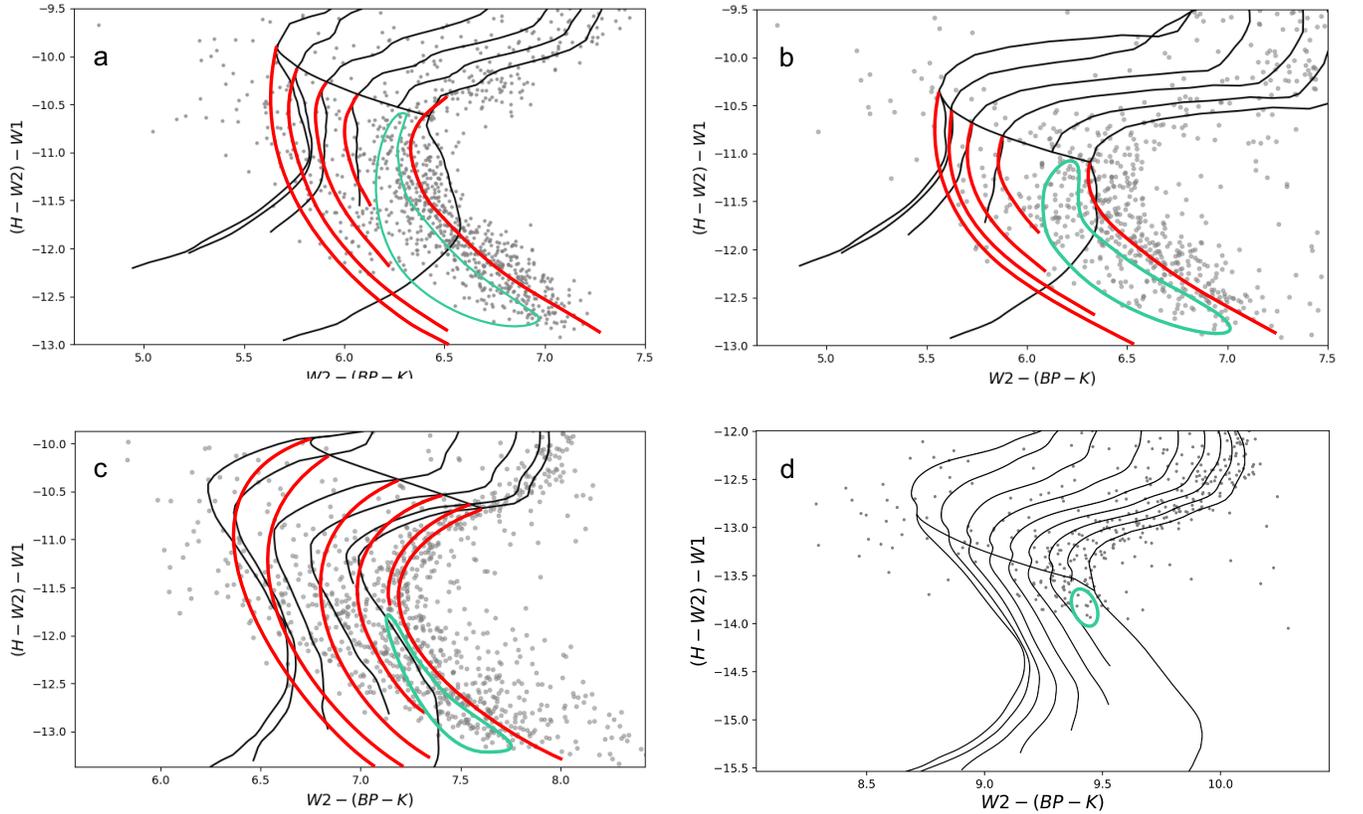}
    \caption{The lower parts of the W2-(BP-K) vs (H-W2)-W1 diagram for all four clusters.
    (a) Pleiades, (b) Alpha Persei, (c) Praesepe, (d) NGC 1039.
    Grey dots indicate cluster stars, while black lines are theoretical lines of constant $q$ values.
    The red lines are the probable `correct' lines of constant $q$ values plotted by eye following the cluster sequence.
    The green lines outline the probable region occupied by binary systems with the secondary component of the mass lower than 0.1 $M_{\odot}$.
    }
     \label{low-mass}
\end{figure}

\noindent
All clusters show a large number of a very low-mass secondary components in the binary systems with primary components below 0.5 $M_{\odot}$.
We expect that in general these secondary components are brown dwarfs.
This point is illustrated in Fig.12 for all four clusters.
The general problem is that the theoretical isochrone (line of $q=0$) and theoretical lines for constant $q$ values do not coincide well with the clusters' sequence, as already stressed.
Due to this fact we tried to position by eye the `correct' $q$ lines following the cluster sequence (red lines in Fig.12).
In the case of NGC 1039 we could not achieve this because the cluster sequence sharply disappears below the primary component for 0.5 $M_{\odot}$ line.
The green lines in Fig.\ref{low-mass} outline the probable region occupied by binary systems with the secondary component mass smaller than 0.1 $M_{\odot}$.
They also were positioned by eye.

Clearly, we cannot consider these arguments as strong evidences.
Most of all, the difficulties of fitting isochrones to the cluster sequence hampers a more solid and quantitative analysis.
Besides, in the case of the faintest stars, we cannot exclude a fraction of random contamination among cluster probable members.
In spite of all these limitations, we believe that the presence of the low-mass components in the binary systems and the brown dwarfs among them is quite probable.

In the considered mass range the binary and multiple star ratio is between $\alpha=0.45\pm0.03$ for Praesepe, and $\alpha=0.73\pm0.03$ for Pleiades.
The ratio of the multiple stars with multiplicity greater than 2 is between $0.06\pm0.01$ for Alpha Persei and Praesepe, and $0.09\pm0.02$ for NGC 1039.
These estimates are larger than previously found (see a review above and in \citet{Malofeeva+2022}).
We can easily account for this result since our diagram allows one to detect unresolved binaries with smaller component mass ratio which were clearly missed in previous investigations.

We provide also a lower estimates for the number of the triple and quadruple systems.
Fig.3, Fig.5, and Fig.6 show the presence of stars which corresponds to the theoretical location of systems with the multiplicity greater than 4.
However, the presence of such systems seems rather unreliable because stable systems with the multiplicity greater than four should be resolved at distances of the Pleiades, Alpha Persei, and Praesepe clusters.
On the other side, it seems that the probability to find stable quintuple or sextuple system in the cluster is very low because such systems should be wide (hierarchic) and easily disrupted by encounters with other stars.
The origin of stars in the far left side of the diagram W2-(BP-K) vs (H-W2)-W1, in fact,  could be explained, for example, by the random presence of field stars in the sample or by large photometric errors.

\noindent
In the future we plan to use the lists of the cluster probable members from \citet{Cantat-Gaudin+2020} to investigate the binary and multiple star population of a larger number of clusters.
Besides, we will try to expand the range of stellar mass for the investigation.

\begin{acknowledgments}
The work of A.F.Seleznev was supported by the Ministry of Science and Higher Education of the Russian Federation, FEUZ-2020-0030. The work of G. Carraro has been supported by Padova University grant BIRD191235/19: {\it Internal dynamics of Galactic star clusters in the Gaia era: binaries, blue stragglers, and their effect in estimating dynamical masses}.
\end{acknowledgments}

\bibliography{Pleiades_ubms_1}{}
\bibliographystyle{aasjournal}



\end{document}